\documentclass[apjl]{emulateapj}
\usepackage{psfig,amsfonts,amsmath,graphicx,natbib,apjfonts}
\def\ra#1#2#3{#1$^{\rm h}$#2$^{\rm m}$#3$^{\rm s}$}
\def\dec#1#2#3{#1$^\circ$#2$'$#3$''$}

\def\swift{{\it Swift}}

\def\grb{GRB\,130603B}
\def\nar{New Astronomy Reviews}

\def\cfa{1}

\begin{document}

\title{An r-Process Kilonova Associated with the Short-Hard
GRB\,130603B}

\author{
E.~Berger\altaffilmark{\cfa},
W.~Fong\altaffilmark{\cfa}, and
R.~Chornock\altaffilmark{\cfa}
}

\altaffiltext{1}{Harvard-Smithsonian Center for Astrophysics, 60
Garden Street, Cambridge, MA 02138, USA}

\begin{abstract} We present ground-based optical and {\it Hubble Space
Telescope} optical and near-IR observations of the short-hard
GRB\,130603B at $z=0.356$, which demonstrate the presence of excess
near-IR emission matching the expected brightness and color of an
r-process powered transient (a ``kilonova'').  The early afterglow
fades rapidly with $\alpha\lesssim -2.6$ at $t\approx 8-32$ hr
post-burst and has a spectral index of $\beta\approx -1.5$
($F_\nu\propto t^\alpha\nu^\beta$), leading to an expected near-IR
brightness at the time of the first {\it HST} observation of $m_{\rm
F160W}({\rm t=9.4\,d})\gtrsim 29.3$ AB mag.  Instead, the detected
source has $m_{\rm F160W}=25.8\pm 0.2$ AB mag, corresponding to a
rest-frame absolute magnitude of $M_J\approx -15.2$ mag.  The upper
limit in the {\it HST} optical observations is $m_{\rm F606W}\gtrsim
27.7$ AB mag ($3\sigma$), indicating an unusually red color of
$V-H\gtrsim 1.9$ mag.  Comparing the observed near-IR luminosity to
theoretical models of kilonovae produced by ejecta from the merger of
an NS-NS or NS-BH binary, we infer an ejecta mass of $M_{\rm
ej}\approx 0.03-0.08$ M$_\odot$ for $v_{\rm ej}\approx 0.1-0.3c$.  The
inferred mass matches the expectations from numerical merger
simulations.  The presence of a kilonova provides the strongest
evidence to date that short GRBs are produced by compact object
mergers, and provides initial insight on the ejected mass and the
primary role that compact object merger may play in the r-process.
Equally important, it demonstrates that gravitational wave sources
detected by Advanced LIGO/Virgo will be accompanied by optical/near-IR
counterparts with unusually red colors, detectable by existing and
upcoming large wide-field facilities (e.g., Pan-STARRS, DECam, Subaru,
LSST).  \end{abstract}


\section{Introduction}
\label{sec:intro}

Over the past decade there has been growing circumstantial evidence
linking short-duration gamma-ray bursts (GRBs) with the coalescence of
compact object binaries (NS-NS and/or NS-BH;
\citealt{elp+89,pac91,npp92,ber11}).  This evidence includes the
location of some short GRBs in elliptical galaxies
\citep{bpc+05,gso+05,bpp+06,fbc+11,fbc+13}; the absence of associated
supernovae \citep{hsg+05,hwf+05,sbk+06,ktr+10}; the distribution of
explosion site offsets relative to the host galaxies, extending to a
distance of $\sim 100$ kpc and matching population synthesis
predictions for compact object binaries \citep{ber10,fbf10,fb13}; and
the lack of spatial correlation between short GRB locations and the
underlying distribution of star formation or stellar mass in their
hosts \citep{fbf10,fb13}.  The combination of these properties clearly
points to the binary merger model, but we currently lack a direct
signature such as the coincident detections of gravitational waves.

Another expected signature of the merger model is an optical/infrared
transient powered by r-process radioactive elements produced by the
ejection of neutron-rich matter during the merger, a so-called
kilonova (e.g., \citealt{lp98,mmd+10,bk13}).  Recent simulations
suggest an ejected mass of $\sim (0.5-5)\times 10^{-2}$ M$_\odot$
(depending on the mass ratio of the binary constituents) with a
velocity of $\sim 0.1-0.3c$ (e.g., \citealt{gbj11,pnr13}).  In
addition, initial calculations by \citet{bk13}, now confirmed by
others \citep{rka+13,th13}, indicate that due to the large opacity of
r-process elements such a transient is expected to peak in the near-IR
with a timescale of $\sim 1$ week and an absolute magnitude of only
$M_J\sim -15$ mag.  In the optical band the timescale is expected to
be shorter, with a strongly suppressed peak brightness (e.g., 3 days
and $\sim -13$ mag in $I$-band; 1 day and $\sim -11$ mag in $B$-band).
Despite their low luminosity and fast timescale, kilonovae are of
great interest as a detectable and isotropic counterpart to
gravitational wave sources from the upcoming Advanced LIGO/Virgo
experiments (e.g., \citealt{mb12}).  At the same time, such transients
should accompany short GRBs, if they can be discerned against the
generally brighter and bluer afterglow emission.

In recent years there have been a few unsuccessful searches for a
kilonova signature in short GRBs \citep{bpp+06,pmg+09,ktr+10}, but
these were focused in the optical band, which current models show to
be strongly suppressed \citep{bk13}.  In this {\it Letter} we present
the first detection of a kilonova, associated with GRB\,130603B at
$z=0.356$.  The results are based on a combination of early
ground-based optical observations, coupled with optical and near-IR
{\it HST} observations at a rest-frame time of about 1 week that
reveal excess near-IR emission with an absolute magnitude and red
optical/near-IR color that closely match the kilonova predictions.
The presence of a kilonova provides strong evidence for compact object
mergers as the progenitors of short GRBs, has crucial implications for
the identification of electromagnetic counterparts to gravitational
wave sources, and is indicative of compact object mergers as the
primary site for the r-process.

\section{Observations and Analysis}
\label{sec:obs}

\begin{figure*} 
\centering
\includegraphics[angle=0,height=4in]{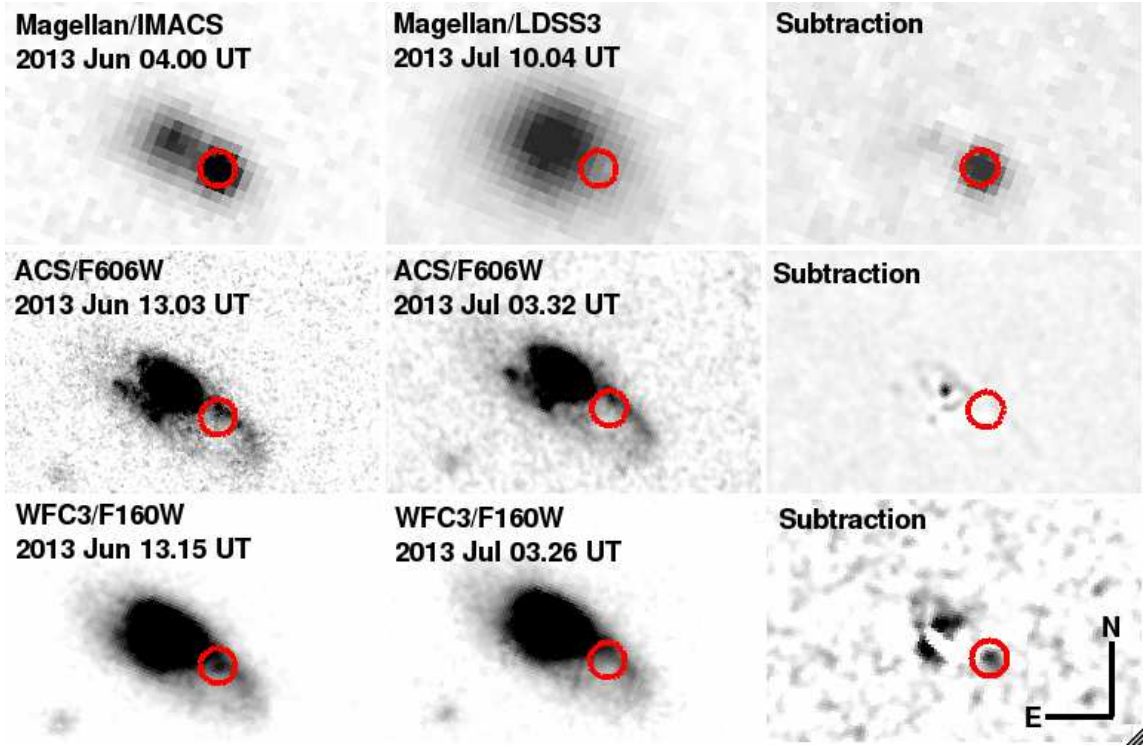}
\caption{Magellan and {\it HST} observations of the afterglow and host
galaxy of \grb.  {\it Top}: Magellan/IMACS and LDSS3 $r$-band
observations at $8.2$ hr and 36.4 d, respectively, with the location
of the afterglow marked. {\it Middle}: {\it HST}/ACS/F606W images.
{\it Bottom}: {\it HST}/WFC3/F160W images.  In all three rows the
circle marks the location of \grb, with a radius of $10$ times the rms
of the astrometric match between the Magellan and {\it HST} images
($1\sigma\approx 34$ mas).  A fading source coincident with the
afterglow position is clearly visible in the WFC3/F160W image, with no
corresponding counterpart in the ACS/F606W image.
\label{fig:image}}
\end{figure*}

\grb\ was discovered with the \swift\ Burst Alert Telescope (BAT) on
2013 June 3.659 UT \citep{GCN14735}, and was also detected with
Konus-Wind \citep{GCN14771}.  The burst duration is $T_{\rm 90}=
0.18\pm 0.02$ s ($15-350$ keV), with a fluence of $F_\gamma=(6.6\pm
0.7)\times 10^{-6}$ erg cm$^{-2}$ ($20-10^4$ keV), and a peak energy
of $E_p=660\pm 100$ keV \citep{GCN14735,GCN14771}.  The spectral lags
are $0.6\pm 0.7$ ms ($15-25$ to $50-100$ keV) and $-2.5\pm 0.7$ ms
($25-50$ to $100-350$ keV), and there is no evidence for extended
emission \citep{GCN14746}.  The combination of these properties
indicates that \grb\ is a short-hard burst.  \swift/X-ray Telescope
(XRT) observations commenced about 59 s after the burst and led to the
identification of a fading source, with a UVOT-enhanced position of
RA=\ra{11}{28}{48.15}, Dec=\dec{$+$17}{04}{16.9} ($1.4''$ radius, 90\%
containment; \citealt{GCN14739}).

Ground-based observations starting at about 2.7 hr revealed a point
source slightly offset from a galaxy visible in Sloan Digital Sky
Survey (SDSS) images
\citep{GCN14742,GCN14743,GCN14745,GCN14747,cpp+13}.  The point source
subsequently faded away indicating that it is the afterglow of
GRB\,130603B \citep{cpp+13}.  Spectroscopy of the host galaxy and
afterglow revealed a common redshift of $z=0.356$
\citep{GCN14744,GCN14745,GCN14747,cpp+13,GCN14757}.  We obtained three
sets of $r$-band observations of \grb\ with the Inamori Magellan Areal
Camera and Spectrograph (IMACS) mounted on the Magellan/Baade $6.5$-m
telescope on June 4.00 ($t=8.2$ hr) and 5.00 UT ($t=32.2$ hr), and at
late time with the Low Dispersion Survey Spectrograph (LDSS3) on the
Magellan/Clay telescope on July 10.04 UT ($t=36.4$ d; Fong et al.~in
preparation).  Using digital image subtraction of the IMACS
observations relative to the LDSS3 template image with the ISIS
software package \citep{ala00} we detect the fading afterglow at a
position of RA=\ra{11}{28}{48.166} and Dec=\dec{$+$17}{04}{18.03},
with an uncertainty of 85 mas determined relative to SDSS.  The
centroid uncertainty in the afterglow position is about 10 mas
(Figure~\ref{fig:image}).  The afterglow has $r=21.56\pm 0.02$ mag at
8.2 hr and $r\gtrsim 24.8$ mag ($3\sigma$) at 32.2 hr (all magnitudes
are in the AB system and corrected for Galactic extinction).

\begin{figure*} 
\centering
\includegraphics[angle=0,height=4.8in]{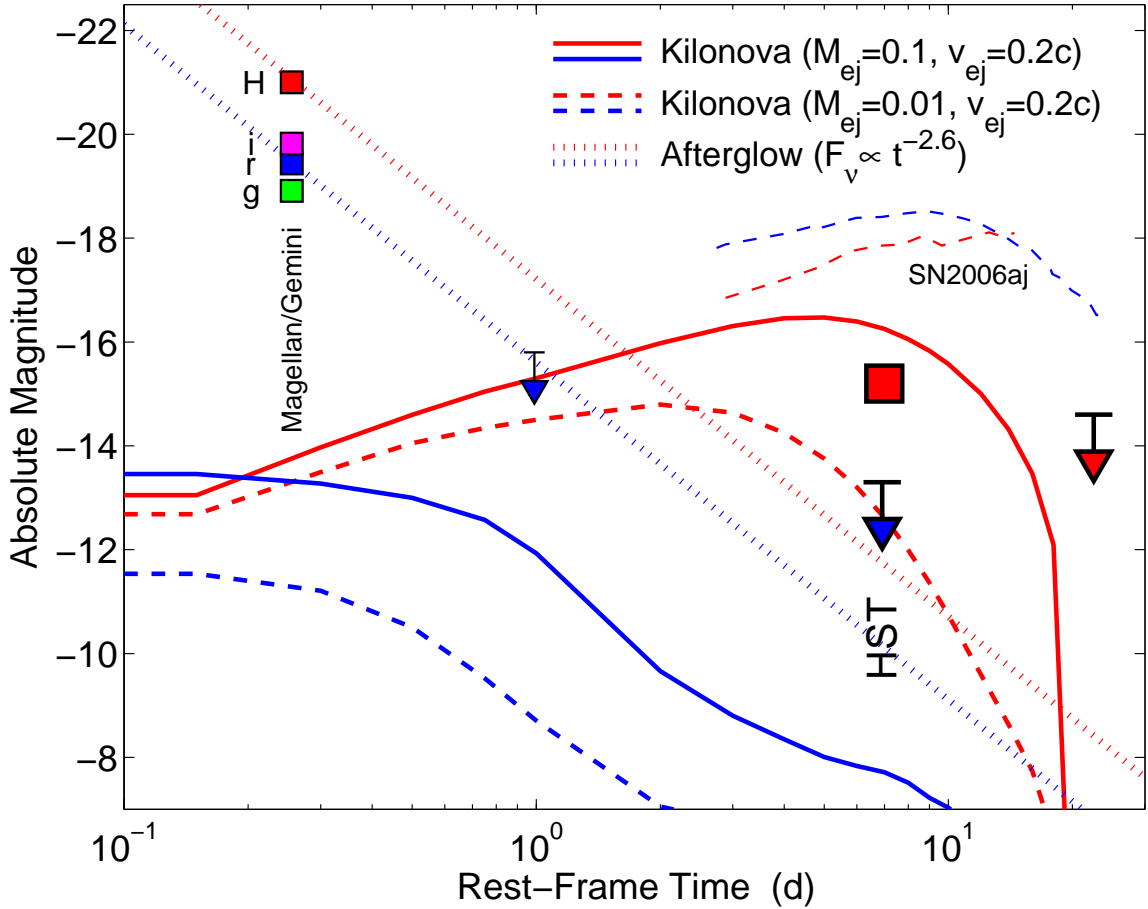}
\caption{Absolute magnitude versus rest-frame time based on our
ground-based observations from Magellan (\S\ref{sec:obs}), on Gemini
data \citep{cpp+13}, and on our {\it HST} photometry (\S\ref{sec:obs};
blue: F606W; red: F160W).  Also shown is an afterglow model with a
single power law decline of $F_\nu\propto t^{-2.6}$, required by the
ground-based observations.  This model underpredicts the WFC3/F160W
detection by about 3.5 mag.  The thick solid and dashed lines are
kilonova model light curves generated from the data in \citet{bk13}
and convolved with the response functions of the ACS/F606W and
WFC3/F160W filters (solid: $M_{\rm ej}=0.1$ M$_\odot$; dashed: $M_{\rm
ej}=0.01$ M$_\odot$).  Finally, we also plot the light curves of
GRB-SN\,2006aj in the same filters (thin dashed;
\citealt{fkz+06,kmb+07}), demonstrating the much fainter emission in
GRB\,130603B, and ruling out the presence of a Type Ic supernova
(\S\ref{sec:kilonova}).
\label{fig:data}} 
\end{figure*}

Two epochs of {\it Hubble Space Telescope} Director's Discretionary
Time observations were undertaken on 2013 June 13.03 UT (ACS/F606W;
2216 s) and 13.15 UT (WFC3/F160W; 2612 s), as well as on 2013 July
3.29 UT (ACS/F606W; 2216 s) and 3.23 UT (WFC3/F160W; 2612 s).  We
retrieved the pre-processed images from the {\it HST} archive, and
distortion-corrected and combined the individual exposures using the
{\tt astrodrizzle} package in PyRAF (Gonzaga et al.~2012).  For the
ACS image we used pixfrac\,=\,1.0 and pixscale\,=\,$0.05$ arcsec
pixel$^{-1}$, while for the WFC3 image we used pixfrac\,=\,1.0 and
pixscale\,=\,0.0642 arcsec pixel$^{-1}$, half of the native pixel
scale.  The final drizzled images, and subtractions of the two epochs
with ISIS, are shown in Figure~\ref{fig:image}.  To locate the
afterglow position on the {\it HST} images, we perform relative
astrometry between the IMACS and {\it HST} observations using 12 and 9
common sources for the WFC3/F160W and ACS/F606W images, respectively.
The resulting rms uncertainty is 34 mas ($1\sigma$).  The subtractions
reveal a fading point source in the WFC3/F160W observations,
coincident with the afterglow position, with no corresponding
counterpart in the ACS/F606W observations (Figure~\ref{fig:image}).

To measure the brightness of the source we use point-spread-function
(PSF) photometry with the standard PSF-fitting routines in the IRAF
{\tt daophot} package.  We model the PSF in each image using 4 bright
stars to a radius of $0.85''$, and apply the WFC3/F160W PSF to the
point source in the subtracted image.  Using the tabulated zeropoint,
we obtain $m_{\rm F160W}=25.8\pm 0.2$ mag.  To determine the limit at
the corresponding position in the ACS/F606W observation, we use the
PSF to add fake sources of varying magnitudes at the afterglow
position with the IRAF {\tt addstar} routine, followed by subtraction
with ISIS, leading to a $3\sigma$ limit of $m_{\rm F606W}\gtrsim 27.7$
mag.  Finally, to obtain a limit on the brightness of the source in
the second epoch of WFC/F160W imaging we add fake sources of varying
magnitudes at the source position and perform aperture photometry in a
$0.15''$ radius aperture and a background annulus immediately
surrounding the position of the source to account for the raised
background level from the host galaxy.  We find a $3\sigma$ limit of
$m_{\rm F160W}\gtrsim 26.4$ mag.  We note that our detection of the
near-IR source was subsequently confirmed by an independent analysis
of the {\it HST} data \citep{tlf+13}.  At the redshift of \grb, the
resulting absolute magnitudes at 9.4 days are $M_H\approx -15.2$ mag
and $M_V\gtrsim -13.3$ mag.

\begin{figure*} 
\centering
\includegraphics[angle=0,height=4.8in]{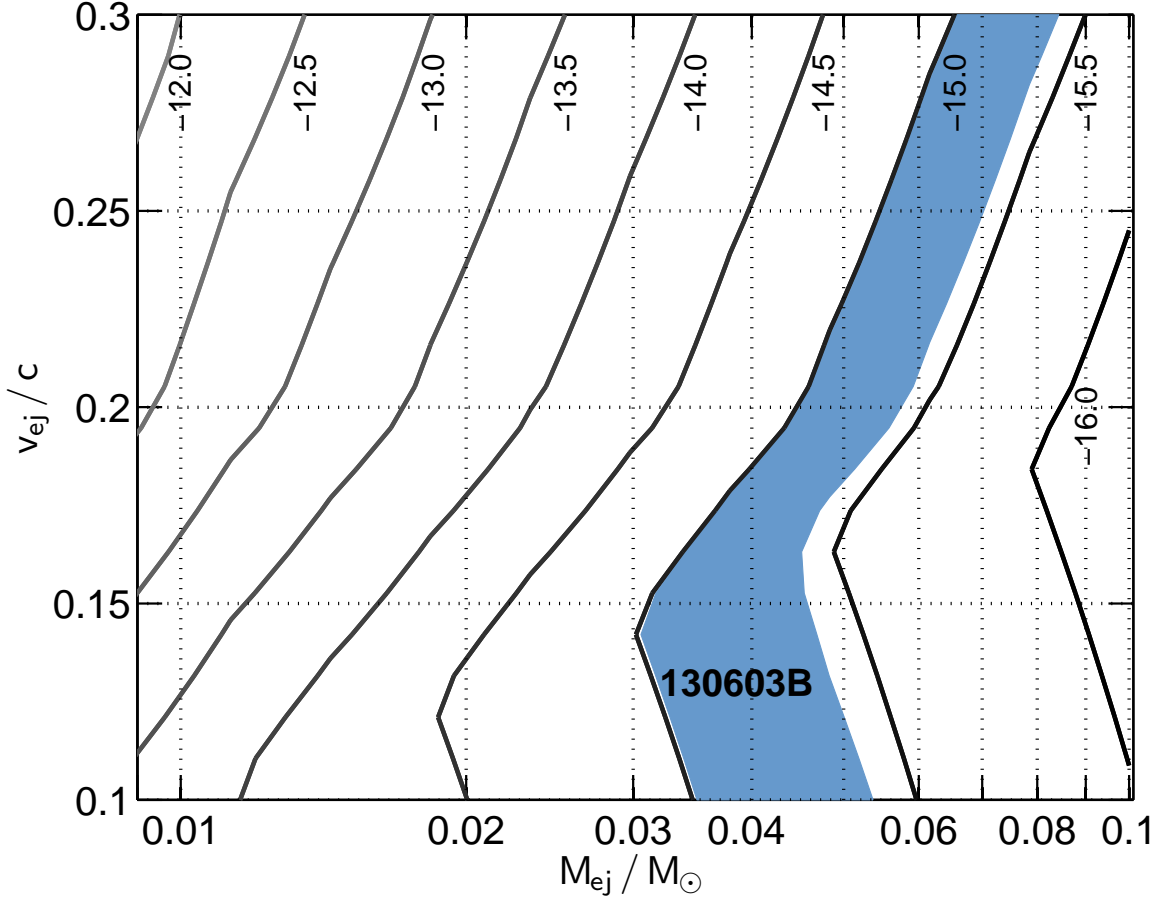}
\caption{Contours of constant kilonova rest-frame $J$-band absolute
magnitude at a rest-frame time of 6.9 days as a function of ejecta
velocity and mass.  The contours were calculated by convolving the
kilonova models from \citet{bk13}, redshifted to $z=0.356$, with the
response functions of the ACS/F606W and WFC3/F160W filters.  The solid
region marks the F160W absolute magnitude of the kilonova associated
with GRB\,130603B.  The observed brightness requires an ejecta mass of
$M_{\rm ej}\approx 0.03-0.08$ M$_\odot$ for $v_{\rm ej}\approx
0.1-0.3c$.
\label{fig:contours}} 
\end{figure*}

\section{An r-Process Kilonova}
\label{sec:kilonova}

In principle, the simplest explanation for the near-IR emission
detected in the {\it HST} data is the fading afterglow.  To assess
this possibility we note that our Magellan optical data at 8.2 and
32.2 hr require a minimum afterglow decline rate of $\alpha\lesssim
-2.2$ ($F_\nu\propto t^\alpha$); $r$-band data from Gemini
\citep{cpp+13} require an even steeper decline of $\alpha\lesssim
-2.6$.  Similarly, the Gemini $gri$-band photometry at 8.4 hr
indicates a spectral index of $\beta\approx -1.5$ \citep{cpp+13},
leading to inferred magnitudes in the {\it HST} filters of $m_{\rm
F606W}\approx 21.6$ mag and $m_{\rm F160W}\approx 20.0$ mag (see
Figure~\ref{fig:data}).  Extrapolating these magnitudes with the
observed decline rate to the time of the first {\it HST} observation
we find expected values of $m_{\rm F606W}\gtrsim 30.9$ mag and $m_{\rm
F160W}\gtrsim 29.3$ mag.  While the inferred afterglow brightness in
F606W is consistent with the observed upper limit, the expected F160W
brightness is at least 3.5 mag fainter than observed.  Moreover, the
afterglow color at 8.4 hr is $m_{\rm F606W}-m_{\rm F160W}\approx 1.6$
mag, while at 9.4 days it is somewhat redder, $m_{\rm F606W}-m_{\rm
F160W}\gtrsim 1.9$ mag, suggestive of a distinct emission component.

The excess near-IR flux at 9.4 days, with a redder color than the
early afterglow, can be explained by emission from an r-process
powered kilonova, subject to the large rest-frame optical opacities of
r-process elements (Figure~\ref{fig:data}).  In the models of
\citet{bk13}, the expected rest-frame $B-J$ color at a rest-frame time
of 7 days (corresponding to the observed ${\rm F606W-F160W}$ color at
9.4 days) is exceedingly red, $B-J\approx 12$ mag, in agreement with
the observed color.  As shown in Figure~\ref{fig:data}, kilonova
models with a fiducial velocity of $v_{\rm ej}=0.2c$ and ejecta masses
of $M_{\rm ej}=0.01-0.1$ M$_\odot$ bracket the observed near-IR
brightness, and agree with the optical non-detection.

In Figure~\ref{fig:contours} we compare the observed F160W absolute
magnitude to a grid of models from \citet{bk13}, calculated in terms
of $M_{\rm ej}$ and $v_{\rm ej}$.  The grid is interpolated from the
fiducial set of models in \citet{bk13}, with $M_{\rm ej}=
10^{-3},\,10^{-2},\,10^{-1}$ M$_\odot$ and $v_{\rm ej}=
0.1c,\,0.2c,\,0.3c$.  We then redshift the models, convolve them with
the response function of the WFC3/F160W filter, and compare the
results to contours of fixed absolute magnitude to determine the model
parameters.  We find that for GRB\,130603B the observed absolute
magnitude requires $M_{\rm ej}=0.03-0.08$ M$_\odot$ for $v_{\rm ej}=
0.1-0.3c$.  Thus, a kilonova can account for the observed emission,
and the observations place a scale of $\sim {\rm few}\times 10^{-2}$
M$_\odot$ for the ejected mass.  We note that using other recent
kilonova models \citep{rka+13,th13} leads to similar results.  The
inferred ejecta mass agrees with the results of merger simulations,
which predict $M_{\rm ej}\sim (0.5-5)\times 10^{-2}$ M$_\odot$
(depending on the mass ratio of the binary constituents;
\citealt{gbj11,pnr13}).

Finally, we stress that the faint emission detected in the {\it HST}
data rules out the presence of an associated Type Ic supernova.  In
particular, at a rest-frame time of $7$ days the GRB-SN\,1998bw had
$M_B\approx -18.2$ mag \citep{gvv+98} at least 4.9 mag brighter than
GRB\,130603B.  Similarly, GRB-SN\,2003lw had $M_J\approx -17.8$ mag
\citep{gmf+04}, about $2.6$ mag brighter than GRB\,130603B.  Even the
relatively dim GRB-SN\,2006aj has $M_B\approx -18.4$ mag
\citep{fkz+06} and $M_J\approx -17.9$ mag \citep{kmb+07}, $\gtrsim
5.1$ mag and $2.7$ mag brighter than GRB\,130603B, respectively
(Figure~\ref{fig:data}).  The normal Type Ic SN\,2002ap had
$M_B\approx -16.5$ mag and $M_J\approx -16.3$ mag on this timescale
\citep{ytk+03}, also well in excess of the observed brightness for
GRB\,130603B.  Indeed, only SN\,2008D had comparable observed absolute
magnitudes, $M_B\approx -13.4$ mag and $M_J\approx -15.6$ mag
\citep{sbp+08,mlb+09}, but this supernova was heavily reddened, with
$E(B-V)\approx 0.6$ mag \citep{sbp+08}.  We therefore conclude that
the short GRB\,130603B was not accompanied by a Type Ic supernova
typical of long GRBs, or even a non-GRB Type Ib/c supernova,
indicating that its progenitor was not a massive star.

\section{Conclusions}
\label{sec:conc}

Using {\it HST} optical and near-IR observations we identify a fading
near-IR source with $M_H\approx -15.2$ mag and $V-H\gtrsim 1.9$ mag at
9.4 days post-burst.  The observed emission is at least 25 times
brighter than expected from an extrapolation of the fading afterglow,
as measured from our Magellan observations and from multi-band Gemini
data \citep{cpp+13}.  A kilonova model with $M_{\rm ej}\approx
0.03-0.08$ M$_\odot$ (for $v_{\rm ej}=0.1-0.3c$) provides a good match
to both the absolute magnitude in the near-IR and the red
optical/near-IR color, making \grb\ the first short burst with
evidence for r-process rich ejecta, a clear signature of compact
object mergers.  The inferred ejecta mass is in good agreement with
the results of numerical simulations for a wide range of compact
object binaries \citep{gbj11,pnr13}.  In addition, the faint
optical/near-IR emission rules out an association with a Type Ic
supernova typical of those that accompany long GRBs, or even non-GRB
Type Ib/c supernovae, demonstrating that the progenitor of
GRB\,130603B was not a massive star.

In addition to providing strong evidence for compact object mergers as
the progenitors of short GRBs, the detection of a kilonova has
additional key implications.  First, the inferred ejecta mass coupled
with the (albeit poorly known) rate of compact object mergers,
suggests that such mergers are likely to be the primary site for the
r-process \citep{ls76,lmr+77,kra+12}.  Second, the observed $H$-band
brightness and the inferred ejecta mass indicate that for a typical
NS-NS merger detected at the Advanced LIGO/Virgo range of 200 Mpc, the
optical $I$-band magnitude will be $\sim 23.5-24.5$ in the first week,
while $J$-band will reach a peak of $\sim 21.5$ mag.  Given the
current lack of wide-field near-IR imagers capable of covering the
typical Advanced LIGO/Virgo localization regions ($\sim 10^2$ deg$^2$)
to this depth, this indicates that searches in the reddest optical
filters ($izy$) with wide-field imagers on large telescopes (e.g.,
Pan-STARRS, DECam, Subaru, LSST) will provide the most promising route
to the electromagnetic counterparts of gravitational wave sources.
\grb\ is likely to become the benchmark for these searches.

\acknowledgments We thank Ryan Foley and Paul Harding for obtaining
the Magellan observations, and Dan Kasen for sharing his kilonova
models.  The Berger GRB group at Harvard is supported by the National
Science Foundation under Grant AST-1107973.  Based on observations
made with the NASA/ESA Hubble Space Telescope, obtained from the Data
Archive at the Space Telescope Science Institute, which is operated by
the Association of Universities for Research in Astronomy, Inc., under
NASA contract NAS 5-26555. These observations are associated with
program \#13497.  This paper includes data gathered with the 6.5 meter
Magellan Telescopes located at Las Campanas Observatory, Chile.

{\it Facilities:} \facility{Magellan}, \facility{HST}


\end{document}